Chiral Magnons: Mechanisms and Research Progress*

Wanxing Lin [1] Hanchen Deng [1] Bao-Tian Wang [2, 3] † Dao-Xin Yao [4, 5, 6] †

1 School of Materials Science and Engineering, Guangdong Ocean University, Yangjiang 529500, China

2 Institute of High Energy Physics, Chinese Academy of Sciences, Beijing 100049, China

3 Spallation Neutron Source Science Center, Dongguan 523803, China

4 School of Physics, Sun Yat-Sen University, Guangzhou 510275, China

5 Guangdong Provincial Key Laboratory of Magnetoelectric Physics and Devices, Guangzhou 510275, China

6 Research Center for Magnetoelectric Physics of Guangdong Province, Guangzhou 510275, China

## Abstract

Chiral magnons are distinctive collective spin excitations in magnetic ordered systems, whose dispersion relations break momentum-inversion symmetry, $\omega(\boldsymbol{k}) \neq \omega(-\boldsymbol{k})$, resulting in essential non-reciprocal spin-wave propagation. This built-in directionality provides new opportunities for spin information transfer, thermal-spin interconversion, and low-dissipation non-reciprocal microwave devices, which complement but differ from topological magnonics. In recent years, the proposal and rapid development of altermagnetism have broadened the physical origin and research framework of chiral magnons, making them a research frontier in condensed matter physics. This review presents a unified framework for chiral magnons, covering symmetry-breaking mechanisms, material implementation, experimental characterization, transport response, and many-body non-Hermitian dynamics, and evaluates routes toward room-temperature and device-related platforms. The discussion is based on symmetry analysis, model Hamiltonians, and spin-wave theory, combined with first-principles calculations as well as recent spectroscopic (e.g., inelastic and polarized neutron scattering, Brillouin light scattering) and transport measurements.

The microscopic origins of chiral magnons can be divided into three interrelated aspects: spin-orbit coupling (SOC)-driven Dzyaloshinskii-Moriya interactions (DMI) in non-centrosymmetric magnets and interfaces; altermagnetism in the weak SOC regime without DMI; the spin space group (SSG) framework. On this basis, representative materials such as CrSb, α-MnTe, α-$Fe_2O_3$, $RuO_2$, and $MnF_2$ are compared in terms of magnetic order and type, physical mechanism, chiral energy scale, coherence, momentum anisotropy, test temperature, and experimental visibility, clarifying how magnon dispersion splitting and lifetimes are reflected in direction-dependent spin Seebeck effects, spin Nernst effects, and thermal Hall signals. At the level of non-reciprocal propagation and device applications, chiral magnons are evolved from intrinsic material properties to artificially engineerable system-level functionalities, thereby paving the way for practical non-reciprocal magnonic devices. This review further summarizes bulk-gap and Berry-curvature induced chiral magnon edge states, the enhancement of non-reciprocity via chiral spin pumping and cavity-magnon hybrids, as well as non-Hermitian features arising from multiparticle damping and gain-loss competition. Besides, remaining challenges, such as the stability of physical properties at room temperature, quantitative calibration of spectral and transport properties, as well as many-body competition, are also outlined. Finally, possible strategies based on SSG-guided material screening, multi-modal metrology, and geometry phase engineering toward efficient spin logic, THz isolators, and quantum routing based on chiral magnons are proposed. This review provides a comprehensive reference for elucidating the underlying mechanisms of chiral magnons, advancing the synthesis and experimental characterization of novel materials, and also guiding the design of next-generation non-reciprocal magnonic devices.

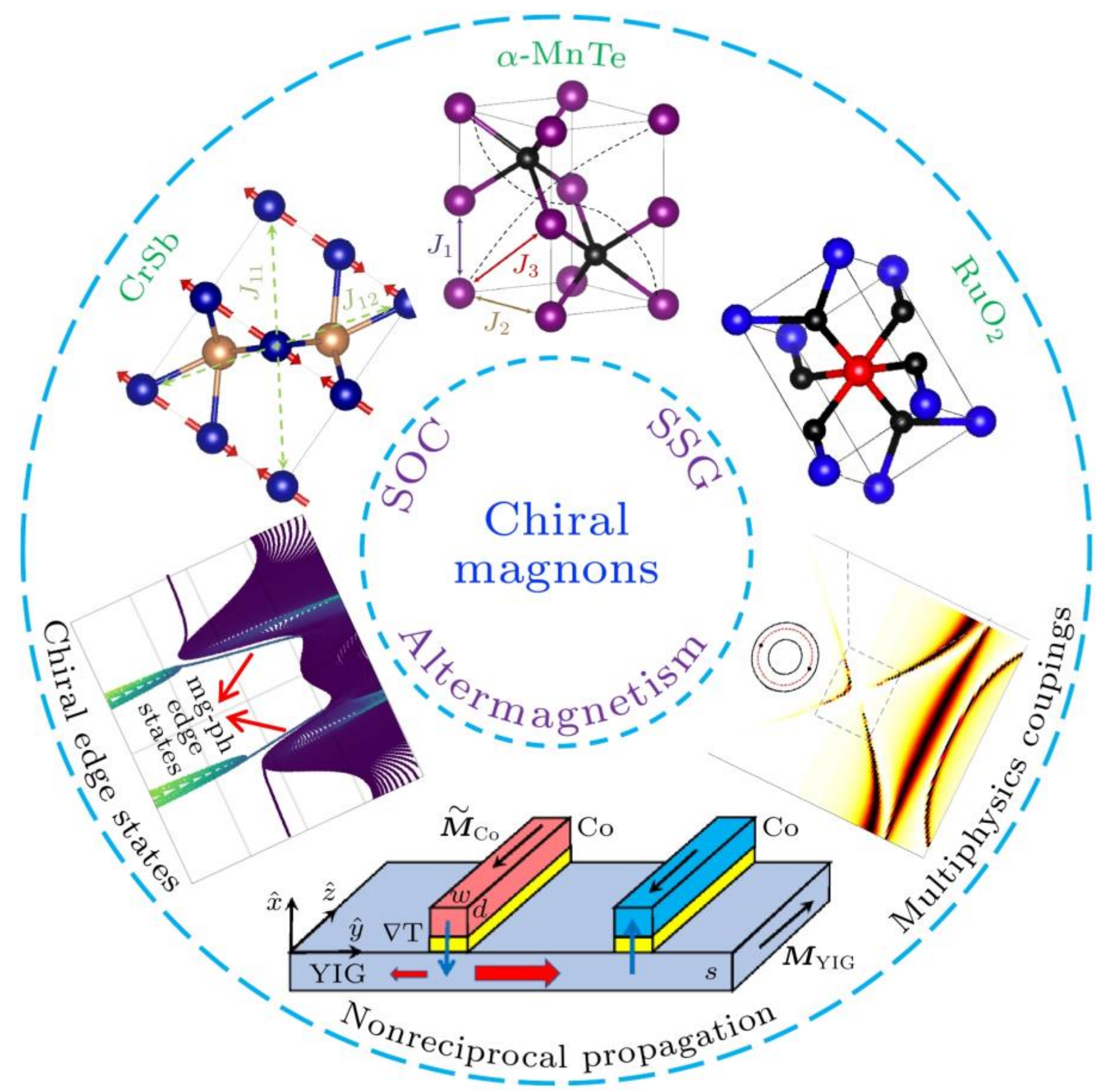

Graphical abstract. Schematic overview of chiral magnons, adapted from Refs. [17, 18, 22, 64, 74, 97].

---

* The paper is an English translated version of the original Chinese paper published in ***Acta Physica Sinica***. Please cite the paper as: W. Lin, H. Deng, B. T. Wang, and D. X. Yao, Chiral magnons: Mechanisms and research progress. *Acta Physica Sinica* **75**: 050706 (2026). doi: 10.7498/aps.75.20251645

Project supported by the National Key Research and Development Program of China (2022YFA1402802); the National Natural Sciences Foundation of China (12304079, 92165204, 12494591, 12574028 and U2330104); The program for scientific research start-up funds of Guangdong Ocean University (YJR22028); the Natural Science Foundation of Guangdong Province (2024A1515010484); Leading Talent Program of Guangdong Special Projects(201626003); Guangdong Provincial Key Laboratory of Magnetoelectric Physics and Devices (2022B1212010008); Research Center for Magnetoelectric Physics of Guangdong Province(2024B0303390001); Guangdong Provincial Quantum Science Strategic Initiative (GDZX2401010).

† Corresponding author. E-mail: wangbt@ihep.ac.cn
Corresponding author. E-mail: yaodaox@mail.sysu.edu.cn
The first author. E-mail: linwx@gdou.edu.cn

## 1 Introduction

Magnons, bosonic quasiparticles that describe collective spin fluctuations in magnetically ordered systems, have been extensively investigated since the establishment of spin-wave theory (SWT) in the 1930s. Research in this field has progressively expanded from low-energy excitations to the broader framework of magnonics, which incorporates topological control and many-body interactions and encompasses dispersion engineering, interaction mechanisms, and strong coupling with light, phonons, and electromagnetic fields[1, 2]. These characteristics endow magnons with unique potential in spin information transfer, non-reciprocal devices, and thermal-spin interconversion. In recent years, the rise of altermagnetism has further directed attention toward chiral magnons-excitations whose asymmetric dispersion arises from the breaking of spatial and spin symmetries rather than from Berry curvature. The defining feature is the momentum-inversion asymmetry $\omega(\boldsymbol{k}) \neq \omega(-\boldsymbol{k})$, which naturally gives rise to non-reciprocal propagation (NRP) and opens parallel application pathways in direction-selective transport and device rectification that are distinct from those of topological magnons[3-11].

Within the overall framework of magnonics, chiral magnons are not an isolated branch but a direct manifestation of symmetry breaking and inequivalent exchange paths. They complement traditional ferromagnetic (FM) and antiferromagnetic (AFM) magnons as well as topological magnons characterized by band topological invariants. Traditional topological magnons primarily rely on Dzyaloshinskii-Moriya interactions (DMI) or the geometric phases of spin textures, and are described by nonzero Berry curvature and Chern numbers. In contrast, chiral magnons can arise from the directional inequivalence of long-range symmetric exchange even in the absence of DMI, producing dispersion splitting $\omega(\mathbf{k}) \neq \omega(-\mathbf{k})$, and further extending non-reciprocity to the odd-symmetric component of the lifetime, $\Gamma(\mathbf{k}) \neq \Gamma(-\mathbf{k})$. This mechanism is particularly prominent in altermagnets. Since its proposal, the concept has rapidly transitioned from theoretical predictions to material screening and experimental verification[12-16]. For example, metallic CrSb has been predicted to exhibit significant chiral dispersion splitting at high energies (~30 meV), although the compression of the observable window by Stoner continuum damping needs evaluation[17]. In the insulator α-MnTe, inelastic neutron scattering (INS) directly observed g-wave type momentum-selective splitting of 2-3 meV, establishing the non-relativistic origin of long-range symmetric exchange[18, 19]. In contrast, INS experiments on $MnF_2$ showed no resolvable splitting, only setting a strict upper limit, revealing that not all AFMs support chiral dispersion[20, 21]. Planar $RuO_2$ further correlates bulk dispersion with in-plane spin Nernst effect (SNE) and spin Seebeck effect (SSE) anisotropy, providing an indirect pathway to infer the splitting amplitude $\Delta\omega(k)$ and magnon lifetime $\Gamma(k)$ from transport measurements[22, 23]. These complementary cases collectively indicate that the observability of chiral magnons depends not only on the splitting amplitude but also on the synergistic constraints of conductivity, spin ordering, and non-Hermitian damping. Currently, high-resolution INS and polarized neutron scattering (PNS) have become core techniques for characterizing chiral magnons. Our experience in neutron scattering phonon dynamics characterization[24, 25] combined with emerging neutron source facilities (such as the China Spallation Neutron Source[26]) is expected to advance magnonics from spectroscopy to comprehensive characterization involving Brillouin light scattering (BLS), THz bidirectional odd-symmetric differential measurements, and neutron-optical polarization[27-30].

In recent years, our research group has continually deepened the work in magnonics, supporting both theoretical and experimental studies of chiral magnons. From indirect K-edge and multi-magnon excitations in triangular and kagome AFMs[31-37], to observations of topological magnon edge states and higher-order Dirac structures[38-43], to torque-balanced SWT with DMI and itinerant spin fluctuations[44-47], these works have not only clarified the role of asymmetric exchange in spectral formation but also provided microscopic images for chiral screening in altermagnets. Recent developments in the field have further extended chiral magnons from intrinsic properties to system functionalities. These include electrically differential detection of left- and right-handed modes in compensated FM-AFM heterostructures[48-50], clarification of constraints imposed by non-local damping on non-reciprocal bandwidth[51, 52], and amplification to GHz broadband isolation using resonant inelastic X-ray scattering (RIXS) circular dichroism momentum imaging and cascaded cavity-magnon platforms[53, 54]. These advances highlight the potential of chiral magnons in multi-physics coupling and non-Hermitian topology. The current research status represents a preliminary integration of theoretical predictions and spectroscopic verification, but breakthroughs are still needed for high-energy-scale room-temperature materials and device integration. Future directions may include strain engineering to optimize inequivalent long-range exchange, development of multi-modal characterization frameworks (e.g., INS-BLS joint measurements), and exploration of unidirectional routing in cavity magnonics to realize prototypes of low-power spin logic and THz isolators.

The research on chiral magnons has formed a parallel framework distinct from the topological mainstream. This review is organized around this framework: Section 2 analyzes its theoretical origins; Section 3 reviews typical material dispersions and non-reciprocal propagation; Section 4 discusses topological and chiral edge states; Section 5 analyzes transport and compensation effects; Section 6 explores many-body non-Hermitian mechanisms; and the conclusion provides an outlook on high-energy-scale room-temperature materials and device directions.

## 2 Theoretical Origins and Symmetry Mechanisms

The core criterion for chiral magnons lies in the momentum-inversion asymmetry of their dispersion relation, i.e., $\omega(\boldsymbol{k}) \neq \omega(-\boldsymbol{k})$. This feature arises from the synergistic action of multiple symmetry-breaking mechanisms. Currently, three complementary lines of inquiry can be summarized: (i) the traditional spin-orbit coupling (SOC)-induced Dzyaloshinskii-Moriya interaction (DMI) mechanism, which provides linear momentum shift and directional selectivity, analogous to the Rashba-Dresselhaus effect; (ii) altermagnetism, which realizes momentum-dependent splitting through non-relativistic exchange in the weak-SOC limit, thereby extending the universality of chiral origins; and (iii) the spin space group (SSG) framework, which unifies the classification of spatial and spin symmetries and provides systematic criteria and material-screening frameworks. These lines progress from microscopic interactions to symmetry protection, not only revealing the physical picture of chiral magnons from degeneracy lifting to non-reciprocal transport, but also providing theoretical support for subsequent material screening, spectroscopic observation, and device design.

In the traditional mechanism, SOC-DMI is the earliest systematically studied source of chiral magnons, particularly in thin films, interfaces, or non-centrosymmetric crystals where spatial inversion symmetry is broken. The DMI term in the Heisenberg model, $\boldsymbol{D}_{ij} \cdot (\boldsymbol{S}_i \times \boldsymbol{S}_j)$, imposes an approximately linear momentum shift, lifting the degeneracy of spin waves in the $\boldsymbol{k}$ and $-\boldsymbol{k}$ directions and producing direction-dependent group velocities[55]. This process is analogous to the spin-momentum locking in the Rashba-Dresselhaus effect in electronic systems. Through interface engineering and anisotropic exchange control of the DMI vector, the non-reciprocal features of magnon dispersion can be artificially constructed in magnetic insulators, as illustrated by the different coupling modes in Fig. 1(a). Furthermore, even in the isotropic Heisenberg model, a small amount of DMI combined with magnon-magnon interactions can open gaps, alter degeneracy points, and regulate dissipation anisotropy, thereby directly determining the non-reciprocal nature of magnon bands and their transport response[56]. As a natural extension of this mechanism, altermagnetism extends chirality to the weak-SOC regime or to zero-SOC, providing an entirely new non-relativistic pathway.

Šmejkal et al. proposed altermagnetism from the perspective of the combined action of lattice symmetry and spin space, marking a transition of collinear magnetic order from traditional FM and AFM toward sublattices related by lattice rotation or point-group mapping. The result is an alternating sign of spin splitting in momentum space, determined by the crystal potential and non-relativistic exchange rather than strong SOC[12]. This mechanism is not only universal for weak-SOC materials but also intersects with spintronics, magnonics, and topological magnets, forming a systematic research landscape (Fig. 1(b)) that provides a conceptual foundation for theoretical predictions and experimental verification[13]. Building on this, the SSG framework further unifies the symmetric treatment of space groups and spin rotations. It classifies collinear magnetic orders into 1421 types (Fig. 1(c)) and screens 498 candidate materials (including 203 altermagnets), thereby providing clear criteria for the protection, lifting of magnon band degeneracies, and g/d-wave angular chiral distribution[56, 57]. In addition to symmetric exchange, interaction effects can also significantly amplify chiral splitting in altermagnets, offering new tunable mechanisms for realizing room-temperature and device-relevant energy scales[58].

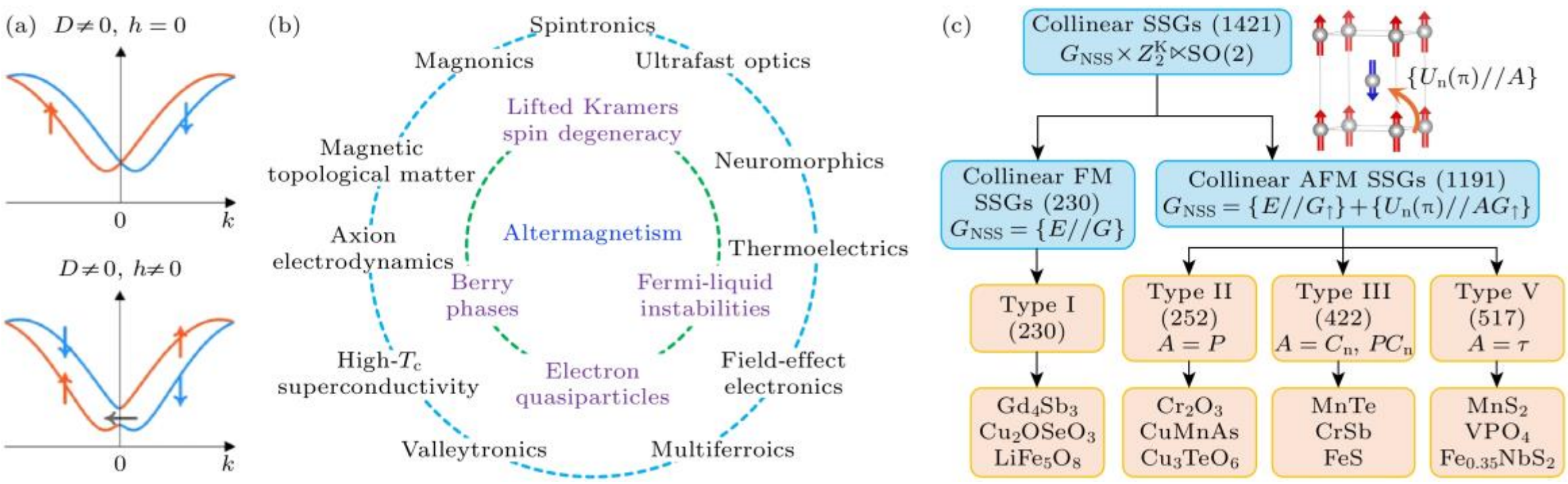

Fig. 1. Theoretical origins and symmetry mechanisms of chiral magnons: (a) Rashba-Dresselhaus analogue for magnons, reproduced with open access from Ref.[55]; (b) research landscape of altermagnetism, reproduced with open access from Ref.[13]; (c) SSG taxonomy for collinear magnets, reproduced with permission from Ref.[56].

## 3 Chiral Magnon Materials and Non-reciprocal Propagation

As a quasiparticle excitation that intrinsically breaks time-reversal symmetry, chiral magnons satisfy $\omega(\boldsymbol{k}) \neq \omega(-\boldsymbol{k})$ in their dispersion relation, and their lifetimes can also exhibit asymmetry, $\Gamma(\boldsymbol{k}) \neq \Gamma(-\boldsymbol{k})$, thereby leading to non-reciprocal propagation. This section focuses on altermagnetic AFMs that do not require DMI. Through systematic comparison of complementary physical pictures in representative materials-metallic CrSb, insulating α-MnTe, α-$Fe_2O_3$, planar metallic $RuO_2$, and the negative control case $MnF_2$, we reveal the energy-scale span of chiral dispersion (from meV to tens of meV), the competition between lifetime and momentum anisotropy, experimental observability windows, and the resulting non-reciprocal transport and device prototypes.

In metallic altermagnets, chiral splitting often reaches higher energy scales but is strongly damped by the Stoner continuum. Taking CrSb as an example, in its A-type collinear AFM ground state, significant chiral splitting appears along the high-symmetry L-Γ-L′ path in the Brillouin zone (BZ). Linear spin-wave theory (LSWT) predicts a splitting of approximately 9 meV, while linear-response time-dependent density functional theory (LR-TDDFT) calculations raise the chiral splitting value to 30-34 meV (Fig. 2), simultaneously revealing Landau damping arising from magnon-Stoner excitation coupling[17]. Many-body perturbation theory further indicates that the observable window of chiral splitting $\Delta\omega(\boldsymbol{k})$ and lifetime $\Gamma(\boldsymbol{k})$ is jointly determined by the competition between chiral exchange and Stoner excitations, providing a theoretical basis for subsequent device-oriented momentum and directional selection[59]. Recent studies on bulk CrSb have also confirmed high-energy chiral splitting along the A-M direction, further validating the coexistence of high energy scale and strong damping in metallic systems[60].

In sharp contrast is the weak-SOC insulating system. Taking α-MnTe as a prototype, its chiral dispersion mainly originates from the inequivalent contributions of long-range symmetric exchange rather than conventional DMI. High-resolution inelastic neutron scattering (INS) directly observed paired splitting of 2-3 meV, with constant-energy surfaces exhibiting six-fold g-wave angular anisotropy (Fig. 3). Theoretical calculations confirm that the splitting remains degenerate in the nodal planes and opens linearly away from them, fully consistent with altermagnetic symmetry constraints[18, 19]. A similar mechanism has been verified in hematite α-$Fe_2O_3$ (Fig. 4(a)). Single-crystal INS measurements give a chiral splitting of approximately 3 meV in the ~100 meV energy region that cannot be explained by conventional AFM exchange models and requires the introduction of higher-order long-range exchange inequivalence. More importantly, this work demonstrates that, under the same crystal symmetry, merely changing the spin ordering type can cause the chiral splitting to disappear or recover, in complete agreement with SSG symmetry criteria[61]. These results collectively establish the central role of long-range symmetric exchange in producing resolvable chiral dispersion in insulators.

Planar metallic $RuO_2$ correlates chiral magnons directly with in-plane spin and thermal transport. Although this system is metallic, THz-scale near-linear chiral splitting still appears near the Γ point (Fig. 4(b)) and exhibits a one-to-one correspondence with the anisotropy of the spin Nernst effect (SNE), spin Seebeck effect (SSE), and thermal Hall effect (Fig. 4(c)). This enables, for the first time, the inference of the directional selectivity of $\Delta\omega(\boldsymbol{k})$ and $\boldsymbol{\Gamma}(\boldsymbol{k})$ through transport measurements[22, 23]. As a negative control case, the classic AFM insulator $MnF_2$ shows only a single spin-wave branch in high-resolution, multi-orientation, and high-energy-resolution INS experiments, strictly symmetric under momentum inversion (Fig. 5). Even under an applied 10 T magnetic field, only Zeeman splitting at the Γ point appears[20, 21]. Polarized neutron scattering (PNS) experiments further suppress the upper limit of the chiral-related signal to an extremely low level[62], strongly proving that, in the non-relativistic limit, a purely collinear AFM structure is insufficient to produce observable chiral dispersion. Altermagnetism combined with inequivalent long-range directional exchange is a necessary condition.

The comparison of the above material systems clearly shows that the energy scale, lifetime, and momentum anisotropy of chiral magnons are not independent parameters but are synergistically regulated by conductivity, SOC strength, and the symmetry of long-range exchange. Currently, high energy scales (>20 meV) mainly appear in metallic altermagnets, but the observation window is compressed by Stoner damping. Low-energy scales (2-3 meV) in insulating systems offer extremely long lifetimes and clear momentum-resolved features, making them particularly well-suited for high-resolution spectroscopic characterization. Planar systems possess the greatest application potential at the transport level. Table 1 summarizes the representative materials and observables. Future breakthrough directions include: (1) precisely tuning long-range exchange inequivalence via strain, doping, or heterostructures to push the splitting energy scale above 10 meV while maintaining long lifetimes; (2) developing joint momentum- and real-space imaging techniques to directly visualize the coupling between g-wave chirality and magnetic domain textures[63].

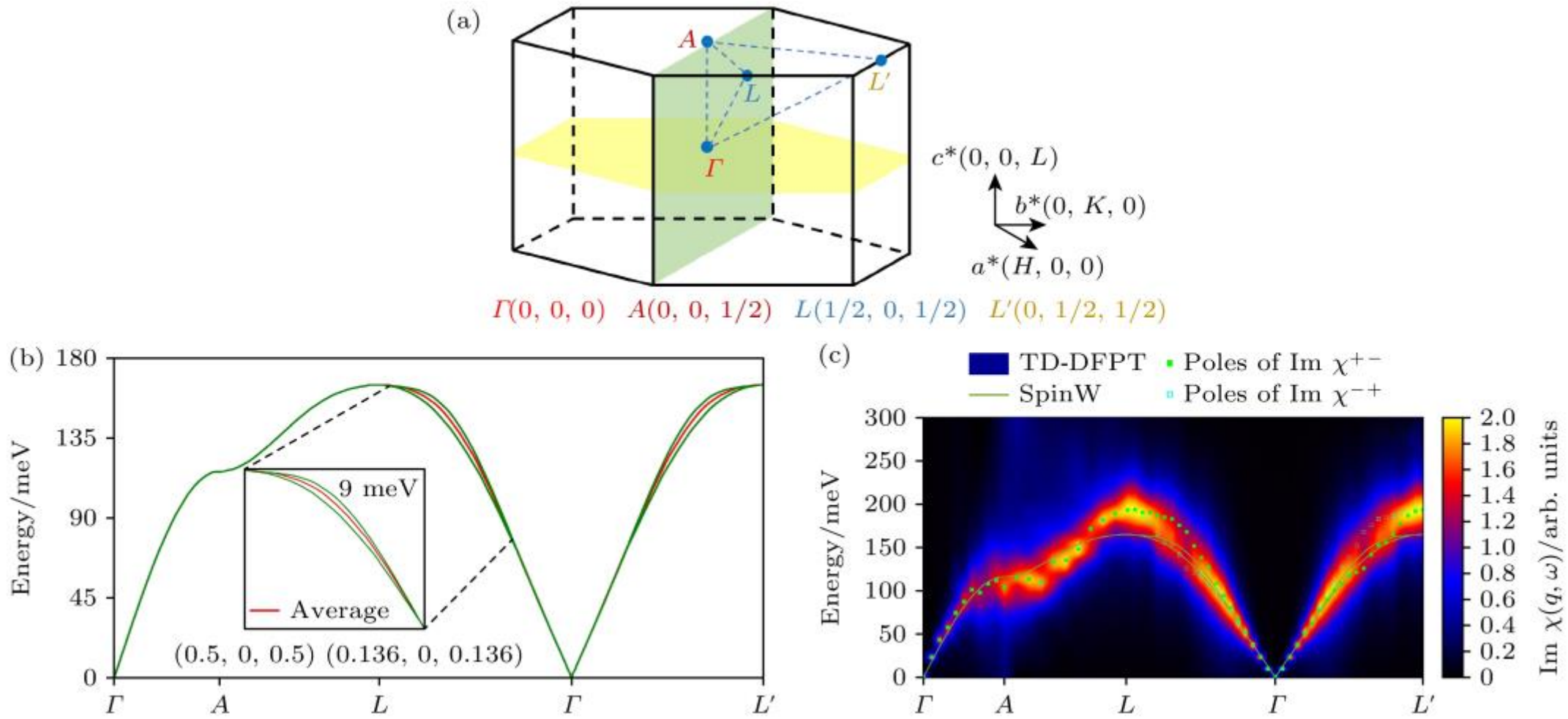

Fig. 2. Chiral magnon splitting in altermagnetic CrSb: (a) BZ and the high-symmetry paths; (b) LSWT magnon dispersion; (c) LR-TDDFT spin fluctuation dispersion (with LSWT overlay) , reproduced with permission from Ref.[17].

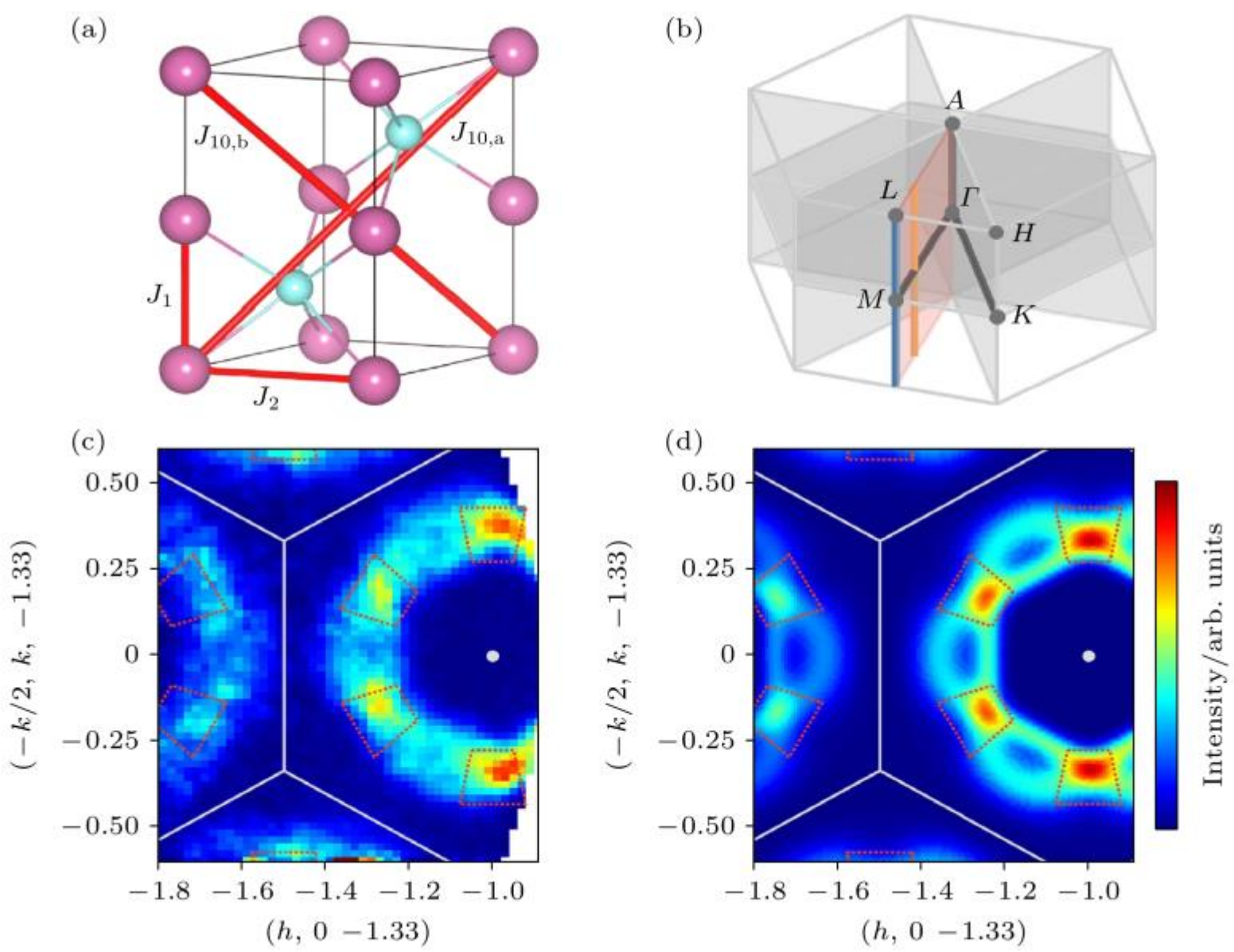

Fig. 3. Symmetric-exchange-origin chiral magnons in α-MnTe. (a) Crystal structure and key exchanges, reproduced with permission from Ref.[19]; (b) BZ and nodal planes; (c) INS constant-energy slices and (d) the corresponding LSWT calculations showing the g-wave six-fold nodes and momentum-selective non-degenerate double branches, reproduced with permission from Ref.[18].

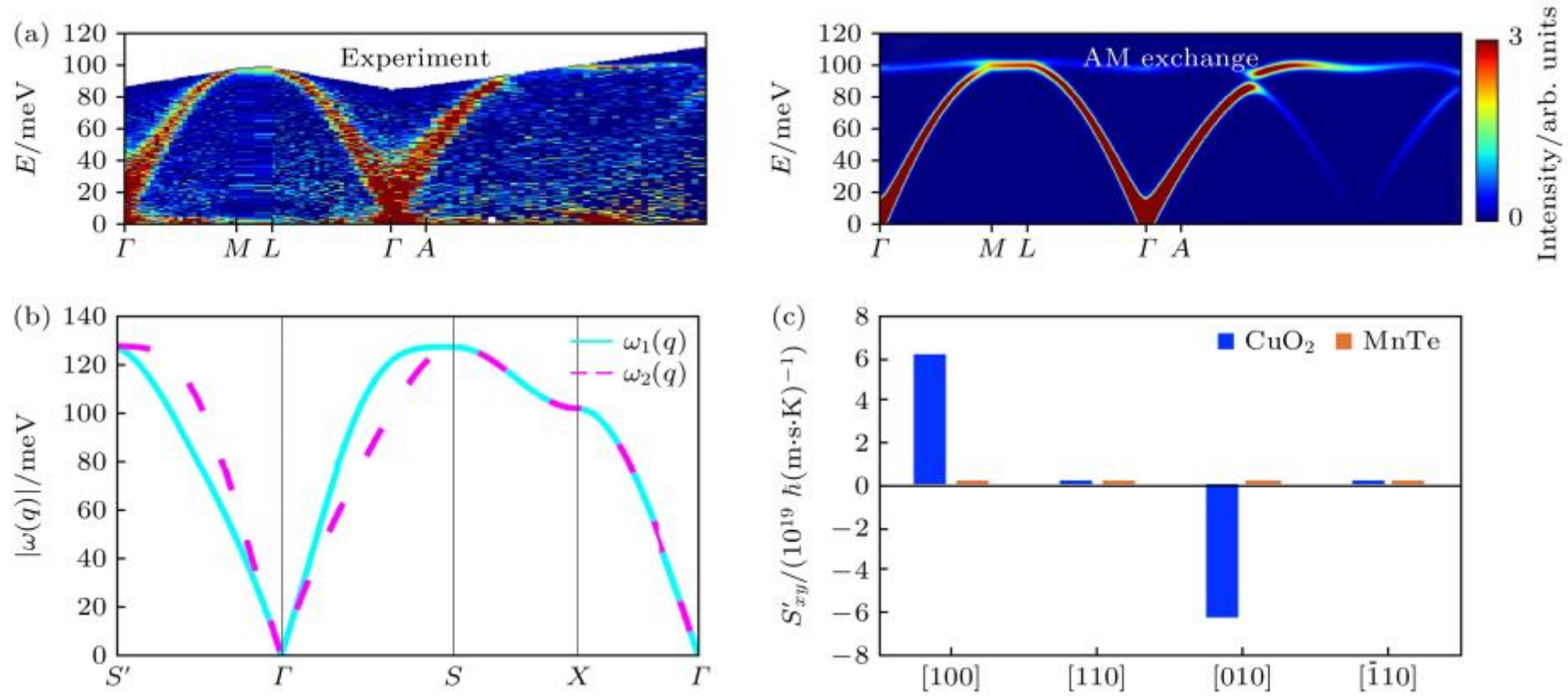

Fig. 4. Chiral magnons of bulk materials and anisotropy: (a) Experimental (the left panel) and calculated (the right panel) INS dispersions of $\alpha$-$Fe_2O_3$ along high-symmetry paths, respectively, reproduced with permission from Ref.[61]; (b) in the $k_x - k_y$ plane, $RuO_2$ shows chiral splitting of magnon modes, reproduced with permission from Ref.[22]; (c) SNE anisotropy of $RuO_2$ vs MnTe, data from Ref.[23].

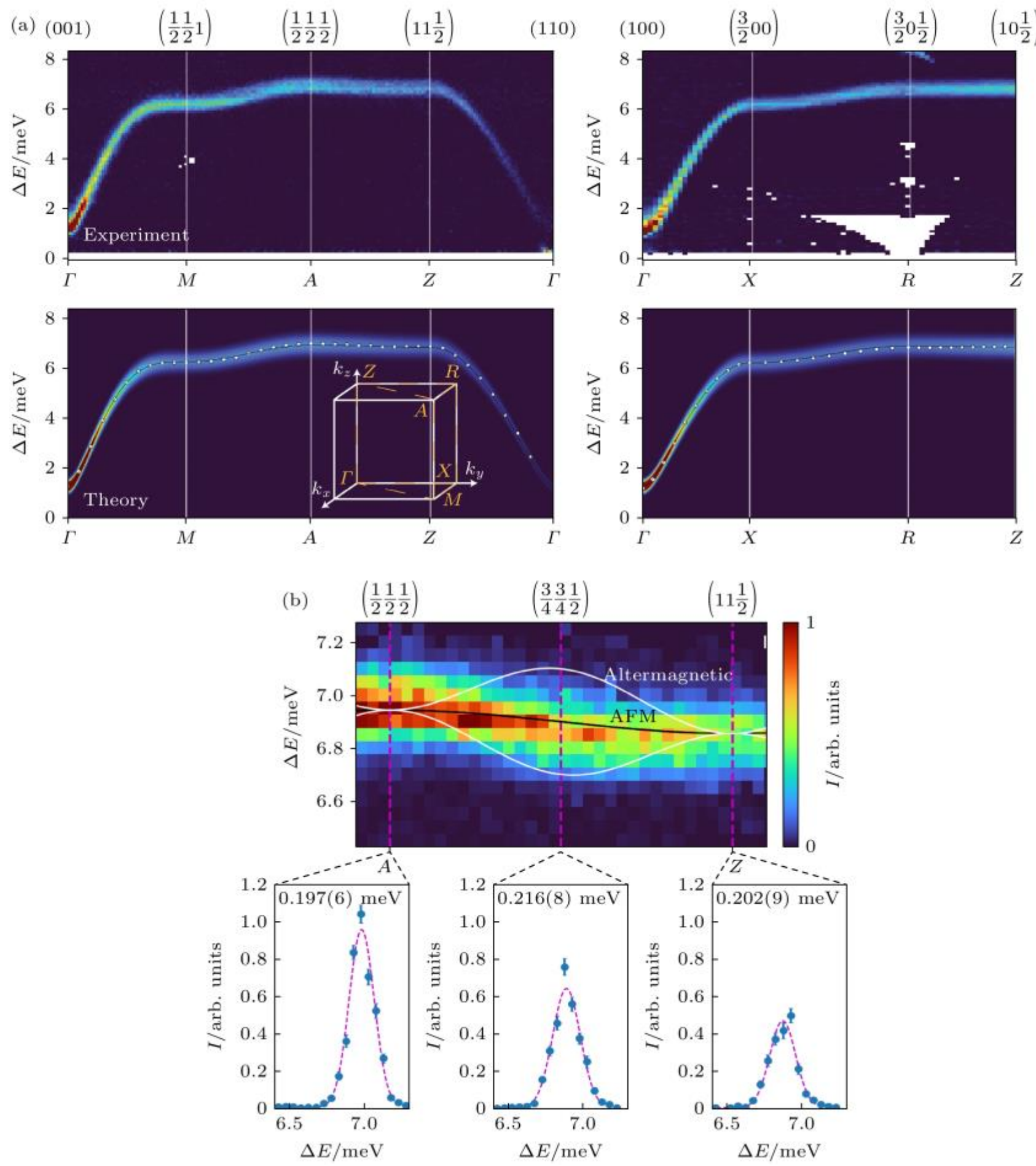

Fig. 5. INS evidence for the absence of chiral magnon splitting in $MnF_2$. (a) (The upper panel) Experimental dispersions and (the lower panel) LSWT fitted dispersions along multiple high-symmetry paths, respectively; (b) high-resolution energy cuts on the ZA line, reproduced with open access from Ref.[21].

Table 1 Representative materials and observables

| System | Magnetic order, type | Main mechanism | Chiral signature (energy scale) | Lifetime, linewidth | Measurement temperature | Observation method |
|---|---|---|---|---|---|---|
| **CrSb** | Metallic altermagnet | Directionally inequivalent long-range exchange | $\Delta\omega \approx 9$ meV (LSWT), $\approx 30$ meV (LR-TDDFT) | High-energy region subject to Landau damping, $\Gamma(\boldsymbol{k})$ anisotropic | 0 K (theory) | LSWT，LR-TDDFT[17] |
| **α-MnTe** | A-type AFM, insulating altermagnet | Inequivalent coupling of long-range exchange | $\Delta\omega \approx 2$ meV; g-wave six-fold angular pattern | Moderate linewidth, directly resolvable by INS | 10 K (experiment); 0 K (theory) | High-resolution INS，LSWT，PNS[18-20] |
| **$RuO_2$** | Metallic altermagnet | In-plane anisotropic exchange and altermagnetic spin symmetry | Near-Γ THz-scale splitting, planar-like morphology | Controllable Landau damping in metallic state, in-plane Γ(k) anisotropic | 0 K (spectroscopic theory); 300 K (transport theory) | Spectroscopic modeling, spin-thermal transport measurements[22, 23] |
| **$MnF_2$** | Collinear AFM | — | No splitting observed | Single spin-wave branch, conventional damping | 1.5 K (experiment); 0 K (theory) | High-resolution INS，PNS[20, 21] |

At the level of non-reciprocal propagation and devices, chiral magnons have evolved from intrinsic material properties to artificially amplifiable system functionalities. Intrinsic non-reciprocity, directly originating from the asymmetric components of dispersion and lifetime, has been mutually verified in the in-plane thermal transport of $RuO_2$ and the momentum-selective splitting of α-MnTe[18, 19, 22, 23]. External engineering provides more flexible control means: chiral spin pumping utilizes the geometric and phase asymmetry of nano-magnet-waveguide structures to selectively enhance the lifetime of a single propagation direction while suppressing the reverse mode (Fig. 6(a)) [27]. Under specific boundary conditions, magnon potential wells can even be formed to achieve directional routing (Fig. 6(b))[64]. Furthermore, long-range non-reciprocal coupling can establish unidirectional effective interactions between spatially separated magnets, forming magnon traps and non-reciprocal transport channels (Fig. 6(c))[65].

Microwave cavity-magnon and magnon-photon strong-coupling platforms amplify the non-reciprocal response to GHz unidirectional passbands and broadband optical isolation[53, 66-68]. These advances collectively indicate that non-reciprocal magnon manipulation has formed a basic framework of material screening, directional excitation, unidirectional transmission, and functional integration.

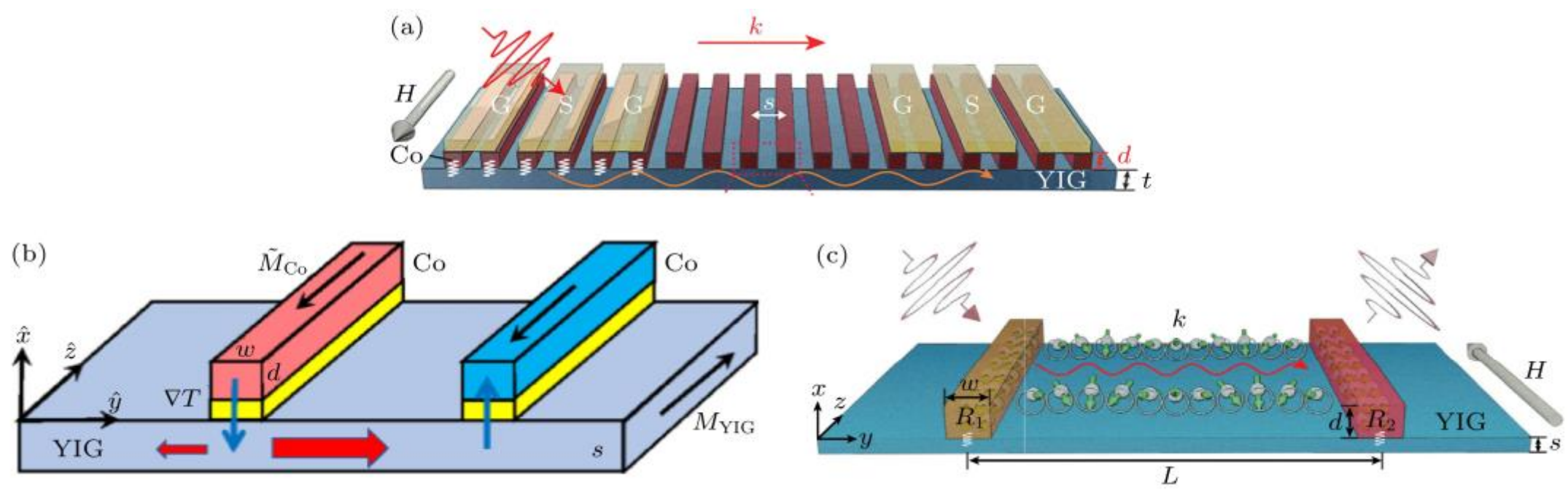

Fig. 6. Engineering the nonreciprocal propagation of chiral magnons: (a) Schematic of the device structure, two coplanar waveguide antennas are fabricated on a Co nanowire grating, which enables unidirectional propagation and chiral injection of spin waves into the yttrium iron garnet film, reproduced with permission from Ref.[27]; (b) schematic of the chiral spin pumping and thermal pumping, reproduced with permission from Ref.[64]; (c) on the yttrium iron garnet film, two magnetic nanowires are magnetized in-plane parallel to the applied magnetic field, reproduced with open access from Ref.[65].

## 4 Topological and Chiral Edge States

The topological branch of chiral magnons originates from bulk gap opening and non-zero Berry curvature that induces non-trivial topological numbers. Even when the bulk dispersion maintains $\omega(\boldsymbol{k}) = \omega(-\boldsymbol{k})$ over most of the momentum range, unidirectional, backscattering-immune edge states can emerge at the boundaries. This causal relationship points from bulk topological indices directly to boundary-localized propagation. It complements the bulk non-reciprocal dispersion and lifetime anisotropy emphasized in Section 3, and the two often reinforce each other directionally in real materials. This section reviews several complementary pathways: exchange-dominated Kagome and honeycomb models provide benchmark frameworks, strong phonon coupling maps chirality onto lattice degrees of freedom, spin textures construct mesoscopic superlattices, Floquet driving injects time-dependent topology, and layered dimensionality introduces phase-transition criticality. These pathways progress from microscopic exchange to multi-degree-of-freedom control, revealing not only the generation mechanisms of edge states but also providing a materials foundation for coherent driving in the GHz frequency range and detection via thermal Hall and spin Nernst effects[69].

In two-dimensional (2D) quantum magnets dominated by exchange interactions, the Kagome ferromagnetic insulator marks an early INS observation of chiral edge states. In this system, the out-of-plane DMI on nearest-neighbor bonds opens a bulk gap for the three magnon bands under an applied polarizing field (Fig. 7(a)), imparting a non-zero Chern number and predicting unidirectional propagating chiral magnon edge states within the bulk gap. This is consistent with the subsequently observed finite thermal Hall signals and is regarded as a representative 2D topological magnon insulator[3], providing an energy-scale reference for subsequent flat-band and Van Hove singularity detection[70]. Further comparison with the honeycomb Kitaev-Heisenberg-Γ-Γ′ model, which replaces DMI with anisotropic exchange, shows that a strong field along the [111] direction introduces an anomalous term that opens a bulk gap. Both LSWT and numerical calculations consistently confirm the unidirectionality of edge states and the variation of their coherence length[71, 72]. Mild doping combined with finite Γ exchange can further induce a dynamical ferromagnetic state, extending chiral features to the synergistic regulation by field, doping, and exchange interactions[73]. These results collectively indicate that exchange inequivalence is not only the source of the bulk gap but also determines the direction of edge propagation and transport reciprocity through the topological charge, suggesting the feasibility of inferring edge states from SNE or thermal Hall conductance characteristics.

As a natural extension of spin degrees of freedom, strong phonon coupling and spin textures further broaden the channels for edge-state generation and detection. In 2D magnets lacking inversion symmetry, circularly polarized phonons and magnons form polarons with selection rules at band crossings. These not only inherit the edge states within the bulk gap of topological magnons but also induce new boundary modes, both exhibit polarization-dependent localized features in the spectrum (Fig. 7(b))[74]. Complementarily, spin textures such as bubbles, vortices, or skyrmion lattices act as periodic superlattices. Localized gyrotropic modes coupled to them simulate the Haldane model, with the edge propagation direction corresponding one-to-one to the topological charge. Simply reversing the skyrmion chirality enables reconfigurable switching (Fig. 7(c))[75-77]. Micromagnetic calculations further reveal the geometric controllability of multiple mirror-image dispersion branches within the bulk gap. These multimodal pathways highlight that edge states are not restricted to the spin channel; acoustic and textural degrees of freedom likewise constitute effective control platforms, thereby enhancing the potential for room-temperature coherence.

Time- and dimensionality-based regulation provide dynamic and critical degrees of freedom that transcend equilibrium states. Floquet driving generates effective magnetic fields and anisotropic exchange under off-resonance conditions, thereby enabling the emergence of chiral edge states even in the absence of a static field. The thermal Hall conductance and Chern number can be continuously tuned with light intensity and polarization, and the target frequency band can be shifted into engineered regimes[78]. In 2D magnon insulators, periodic modulation of DMI can induce topological phase transitions and stable edge states; adjusting the relative phase of the driving can effectively control the chirality of these edge states, offering significant potential for applications in multi-magnon systems[79]. Ferroelectric altermagnets exhibit dual switchable characteristics under ferroelectric polarization, enabling directional control of spin-wave modes and band splitting, thereby introducing regulation of topological edge states. The anticrossing and direction-selective splitting of chiral magnon dispersion generate topology-related edge-state behavior, providing new ideas for edge-state research in topological materials[80]. Weak interlayer coupling in bilayer honeycomb ferromagnets introduces Dirac nodal rings protected by time and space symmetries, moderate DMI allows their coexistence with edge states[81]. Gradually increasing interlayer exchange can trigger the closure and reopening of the bulk gap, accompanied by a reversal of the Berry curvature and Chern number transitions. The thermal Hall coefficient exhibits a sharp peak response at the critical point, serving as a sensitive indicator for identifying phase transitions and boundary reconstruction[82]. Currently, preliminary progress has been achieved in coherent driving and low-damping detection in the GHz frequency range. However, a unified experimental platform is still needed to achieve comprehensive analysis of topological and chiral effects in room-temperature devices through cross-validation of neutron scattering, spectroscopic techniques, thermal Hall effect, and SNE.

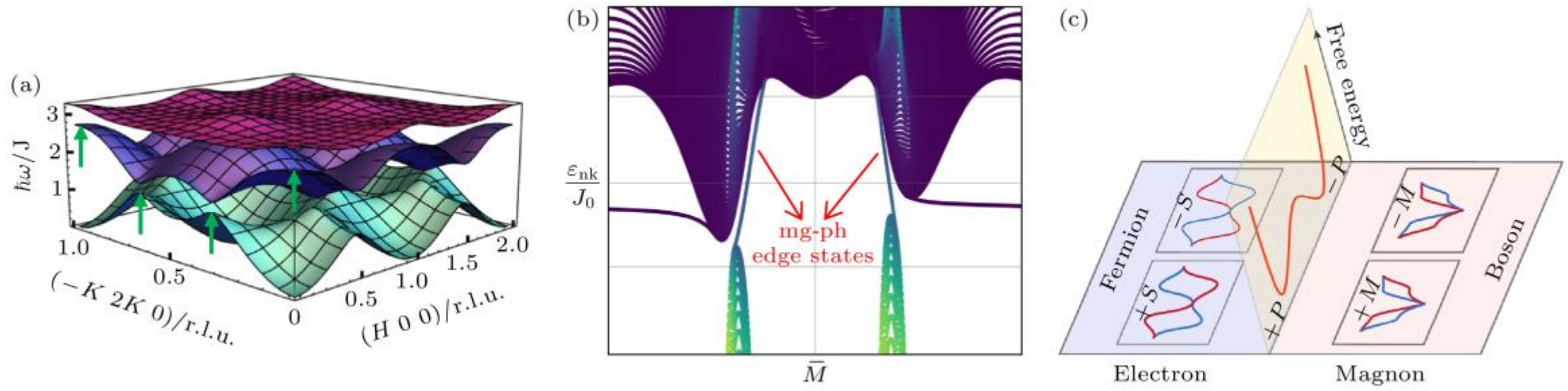

Fig.7. Chiral edge states and topological magnons: (a) Topological magnon bands in a kagome ferromagnet, reproduced with permission from Ref.[3]; (b) magnon-phonon polaron edge states in a Heisenberg-Kitaev magnet, reproduced with permission from Ref.[74]; (c) schematic diagram of magnon chirality splitting and topological edge state regulation induced by electric field reversal in ferroelectric altermagnets, reproduced with permission from Ref.[80].

## 5 Transport and Compensation Effects

The mapping of the non-reciprocal dispersion and damping anisotropy of chiral magnons into macroscopic transport not only provides comparable benchmarks for spectroscopic results but also offers crystal-orientation-sensitive tensor references for device design. Temperature gradients or spin chemical potentials directly introduce the momentum selectivity of $\Delta\omega(\boldsymbol{k})$ and $\Gamma(\boldsymbol{k})$ into the SSE, SNE, and thermal Hall responses. These responses are synergistically constrained by inter-mode competition, interface transparency, and momentum coverage. Consequently, signals can be amplified to room-temperature measurability in strongly chiral metals, easily suppressed in weakly chiral insulators, and exhibit compensation zeros in multi-mode ferrites[28]. This section reviews three complementary threads: SSE and SNE anisotropy evidence in bulk altermagnets establishes the benchmark mechanism, electrically differential transmission in heterolayered structures extends to controllable detection, and multi-mode competition together with emerging RIXS/BLS/neutron platforms mutually verifies the folding effect of damping on signals. These pathways extend from intrinsic system responses to system-level engineering, establishing not only quantitative correspondence with spectra, lifetimes, and transport but also operational pathways for GHz non-reciprocal routing.

In bulk altermagnets, the anisotropy of non-relativistic symmetric exchange simultaneously induces SSE and SNE even under zero external field and weak SOC. Theoretical predictions indicate that the SNE conductance can exceed that of similar AFM monolayers by nearly two orders of magnitude (Fig. 8(a))[83]. Metallic $RuO_2$ exhibits significant tensor sign reversal on the (001) and (110) crystal planes at 300 K, highly consistent with the planar magnon splitting morphology near the Γ point[23]. As a control, although A-type α-MnTe shows ~2 meV momentum-selective splitting in INS, its SNE signal approaches zero[18, 19, 23]. These studies demonstrate that only when $\Delta\omega(\boldsymbol{k})$ is significant over a wide momentum range, and $\Gamma(\boldsymbol{k})$ exhibits clear angular differences, can the transport components be effectively amplified to room-temperature measurable levels. Furthermore, crystal chirality can characterize the orbital angular momentum of magnons; temperature gradients can thereby induce finite orbital responses, analogous to the Edelstein effect in electrons, providing possibilities for orbitalized chiral transport in acoustically coupled systems[84]. This mechanism extends bulk responses from spin-dominated to multi-degree-of-freedom regimes, strengthening the intrinsic connection between spectroscopy and transport.

As a natural extension of bulk systems, heterostructures further introduce chirality into electrically controllable regimes to enhance the resolution of differential detection. In ferrimagnetic-antiferromagnetic layered configurations, different transmission coefficients and phases are imparted to left- and right-handed modes. Through interlayer exchange and interface regulation, non-reciprocal transmission phase engineering is realized[49, 50]. In compensated ferrimagnetic insulators, current excitation combined with voltage detection can directly distinguish analogous chiral signals[85]. In Py/Gd multilayers, ferromagnetic resonance and exchange modes correspond to right- and left-handed precession, respectively. The polarity of spin pumping is determined by the handedness of precession rather than equilibrium magnetization. Sweeping the field enables cyclic switching at fixed frequency and stable detection at room temperature (Fig. 8(b)). This provides electrical evidence for judging the sign and amplitude of chirality-dominated tensors[86]. These advances highlight the bridging role of heterostructure engineering in converting spectroscopic $\Delta\omega(\mathrm{k})$ into device-level routing.

Multi-mode competition reveals the non-monotonicity of transport signals and further requires cross-verification with emerging platforms to decouple damping effects.

In ferrites, for example, redistribution of spin polarization among branches with temperature or field can lead to net spin-current compensation points, essentially reflecting dynamic weighting rather than band crossing[28]. Ignoring this competition easily leads to misjudgment of missing chirality. Complementarily, the degree of circular polarization (DCP) has been used in layered AFMs to resolve the chiral weighting of magnon-phonon coupling, enhancing directional sensitivity of polarization selection (Fig. 8(c))[29]. Emerging techniques such as RIXS circular dichroism momentum imaging of altermagnetic branches[54], scanning transmission electron microscopy combined with meV-level electron energy-loss spectroscopy for NiO nano-THz spectra[87], and neutron spin-echo for μeV-lifetime full-BZ coverage[88], jointly provide element- and space-resolved cross-verification. In $TbMnO_3$, time-of-flight polarized neutrons resolve dynamic chirality flipped by external electric fields, frustrated exchange and anisotropy reproduce the spectral distribution without requiring significant DMI[89]. Recent BLS methodology further decouples DMI (peak-position difference) and chiral damping (odd-symmetric component of full width at half maximum) within the same spectrum, quantifying the odd component of $\Gamma(k)$[90]. This injects odd-even decomposition of damping into the calibration of crystal-orientation dependence of SNE and SSE. Currently, quantitative spectral-transport correspondence has been preliminarily established. Cross-platform combined use (e.g., INS-BLS-RIXS) is anticipated to synchronously calibrate $\Delta\omega(\boldsymbol{k})$, $\Gamma(\boldsymbol{k})$, and compensation zeros under room-temperature wide-momentum windows, enabling prototype validation of low-loss chiral logic.

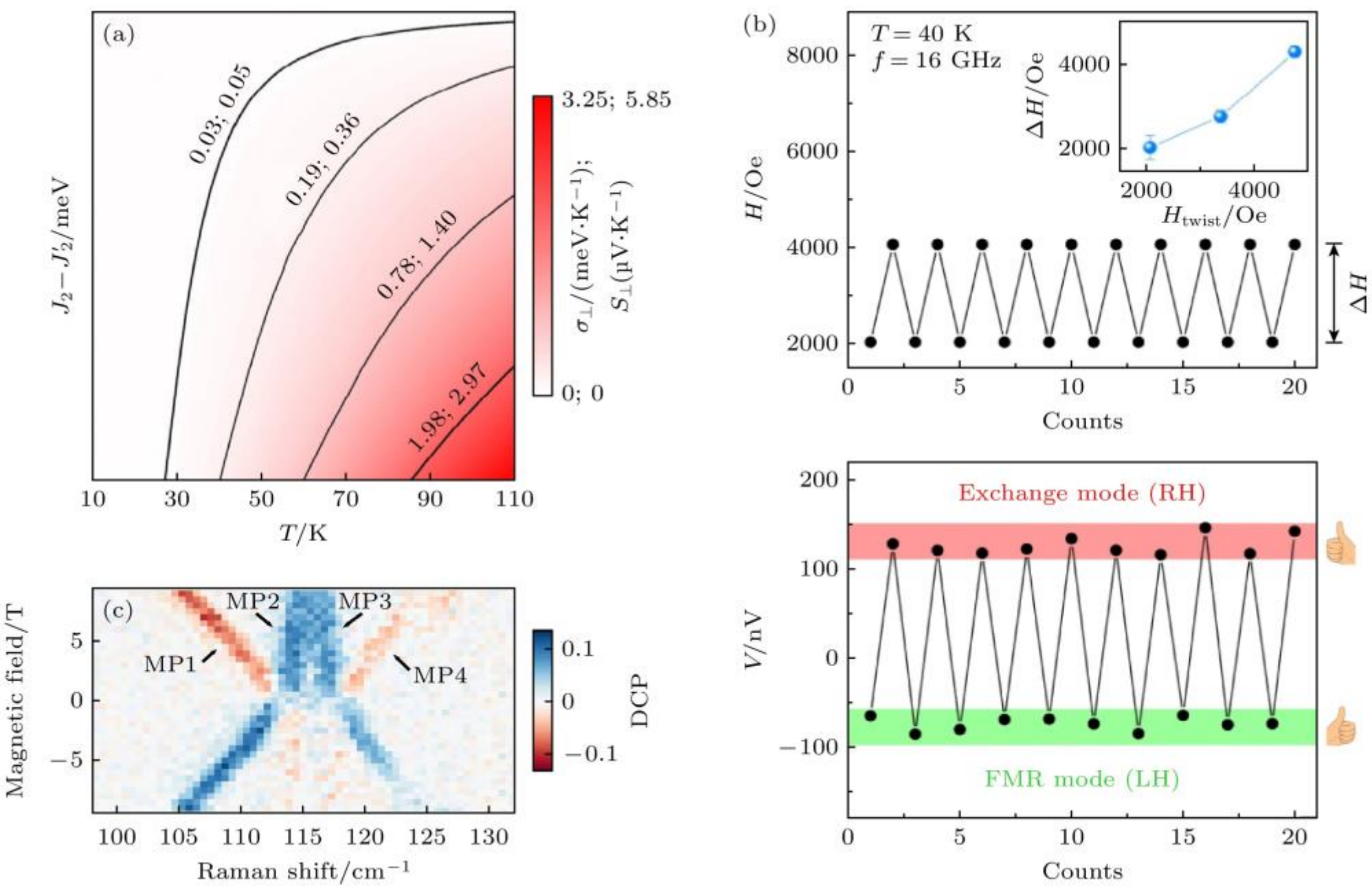

Fig. 8. (a) SNE coefficients depend on temperature and $J_2 - J_2'$, reproduced with permission from Ref.[83]; (b) modulation for chirality of magnon and the spin pumping voltage $V(H)$ signal by recurrent switching of external field $H$, reproduced with open access from Ref.[86]; (c) chiral-selective magnon-phonon mixing in layered zigzag AFM shows a significant contrast with DCP, reproduced with open access from Ref.[29].

## 6 Many-body and Non-Hermitian Aspects of Chiral Magnons

The evolution of chiral magnons in many-body interactions and non-Hermitian environments progresses from condensate coherence to system-level non-reciprocal amplification and further to complex-frequency topological effects, forming a layered regulatory spectrum. Finite-momentum valleys and chiral splitting jointly constrain the threshold and scattering channels of magnon Bose-Einstein condensation (mBEC), suppressing unfavorable processes to expand the stable region. Cavity geometry and phase conditions selectively introduce direction-sensitive coupling, pushing the hybrid response of magnons with photons and mechanical modes toward broadband gain. Non-Hermitian mechanisms further introduce encirclement of exceptional points and skin effects, providing a detection framework for complex-frequency trajectories[91-95]. These pathways use material energy scales (Table 2) as reference and realize cross-scale correlations through Floquet driving, cascaded ring cavities, and active Barnett rotation. They not only map the momentum selectivity of bulk dispersion onto coherent lifetime and transmission coefficients but also provide dynamical tunability for GHz non-reciprocal devices.

In the many-body limit, the synergistic action of chiral splitting and three-magnon/four-magnon scattering dominates mBEC coherence. Moderate chiral splitting suppresses unfavorable channels and expands the stable region, whereas excessively weak or strong splitting leads to mode mixing or high-energy occupation decay, respectively. Accordingly, stability phase diagrams are drawn in the splitting-interaction plane (Fig. 9(a)), providing reference for pump power and frequency-band selection[96]. As a temporal extension of this mechanism, off-resonant Floquet driving generates effective anisotropic exchange and effective fields in the absence of a static field, allowing edge states and thermal Hall responses to be continuously tuned by amplitude and polarization. This facilitates both directional excitation and coherent locking[78]. These results collectively reveal that material-level chiral dispersion presets the energy and momentum windows of coherent states, while external driving determines entry paths and maintenance strategies, further extending many-body coherence from intrinsic system stability to dynamic engineering.

Cavity magnonics elevates directional selection to the system-geometry level, providing an amplification pathway complementary to many-body coherence. Ring and traveling-wave cavities regulate coupling strength via sample radial position and phase, bright and dark mode separation and direction-sensitive transmission appear near anticrossings (Fig. 9(b))[97]. Multi-stage cascading further broadens the non-reciprocal bandwidth[53]. The introduction of mechanical modes into tri-mode coherence produces significant differences between $S_{21}$ and $S_{12}$ within the hybrid band, set by relative phase and strength[98]. In active platforms, Barnett rotation injects different oscillation thresholds into the two modes of relative wavevector, realizing statistical non-reciprocal blocking[99]. Correspondingly, external-field tuning of the band-edge matching barrier parameters in antiferromagnetic $MnF_2$ nanowires produces near-unity transmission resonant ridges in the T coefficient, highly selective to wavevector chirality (Fig. 9(c))[100]. This geometric-phase framework highlights how system conditions determine gain accessibility and controllability, converting the tuning of bulk dispersion into coherent penetration directly observable in the time domain. Recently, in d-wave altermagnets, dipole-dipole interactions have been predicted to realize strong coupling between left- and right-handed exchange magnons at finite wavevectors, manifested as directionally anisotropic band anticrossings and chirality-selective energy gaps[101]. This mechanism injects new pathways for short-wavelength, direction-tunable quantum magnon information processing on altermagnet platforms.

When gain and loss coexist, the non-Hermitian framework extends the spectral function to the complex plane. At exceptional points, eigenvalues and eigenvectors coalesce; opposite encirclements in parameter space produce different complex-frequency trajectories and occupation flips[91-93]. The generalized BZ unifies skin effects and topological classification[94], employing encirclement chirality for non-Hermitian phase identification and sensitive detection[95]. Combined with the aforementioned cavities and tri-mode systems, this mechanism provides explicit criteria for distinguishing whether directional selection originates from symmetric exchange, geometry, or non-Hermitian effects. Currently, many-body coherence and cavity non-reciprocity have been preliminarily realized in GHz prototypes. Cross-use of complex-frequency spectroscopy and Floquet driving is anticipated to quantify the constraints of exceptional points on skin modes over wide temperature ranges, supporting stable integration of room-temperature chiral routing.

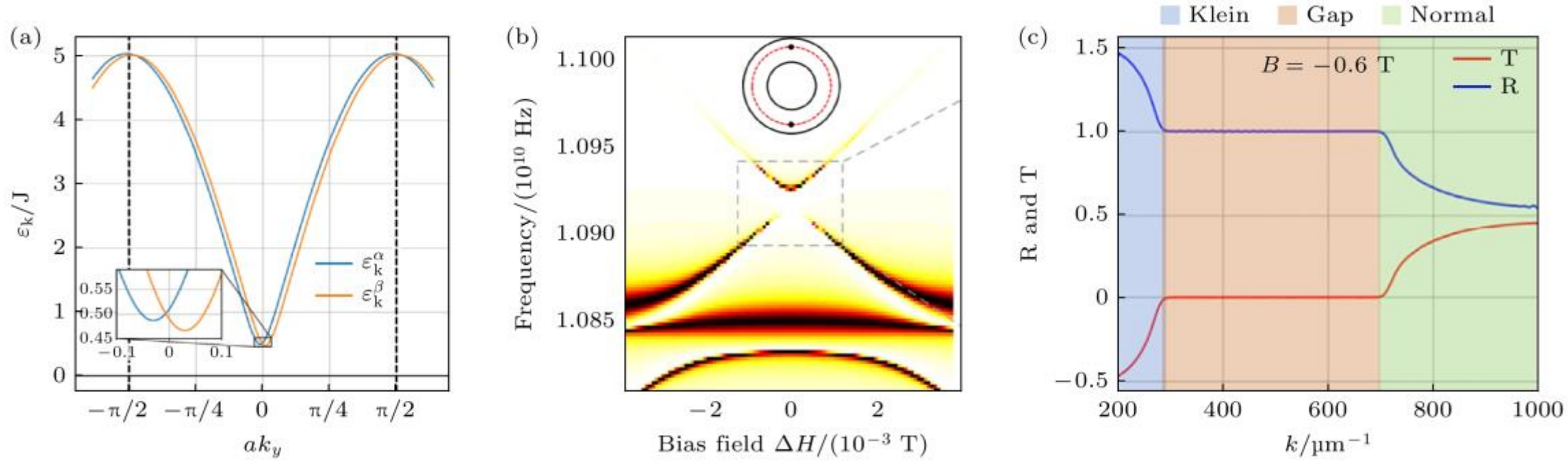

Fig. 9. Multiphysics couplings and non-Hermitian effects. (a) Dispersions and stability domains for right-/left-handed magnons, reproduced with permission from Ref.[96]; (b) magnon-photon strong coupling and bright-dark mode switching in a cavity, reproduced with permission from Ref.[97]; resonant Klein tunneling of chiral magnons in an AFM $MnF_2$ nanowire, reproduced with permission from Ref.[100].

Table 2 Overview of multibody and non-Hermitian platform

| Platform/ Mechanism | Key physics | Tunable parameters | Main observables | Key characterization quantities | Typical energy scale |
|---|---|---|---|---|---|
| **mBEC stability** | Finite ***k*** splitting suppresses unfavorable nonlinear scattering, expands coherent region | Chiral splitting amplitude, interaction strength, pump power | Critical density, lifetime, gain threshold | Two-branch occupation imbalance and stable-region expansion | Material- and pump-dependent[96] |
| **Floquet writing of chiral topology** | Off-resonant driving induces effective field and Berry curvature | Light intensity, polarization, frequency, phase | Edge-mode dispersion, tunable Hall characteristics | Appearance and reversible tuning of chiral edge states | GHz-THz (material-dependent)[78] |
| **Cavity magnon-photon coupling** | Strong coupling and geometric selection lead to directional coupling | Sample radial position, phase | $\|S_{21}\|^2$ transmission map, anticrossing gap | Bright-dark mode switching, direction-selective coupling | GHz-scale Rabi splitting[97] |
| **Cascaded cavity-magnon isolation** | Multi-stage cascading amplifies non-reciprocal bandwidth | Cascade stages, phase matching, coupling strength | $S_{21}$ vs $S_{12}$ bandwidth and suppression | Broadband unidirectional transmission | GHz[53] |
| **Magnon-phonon coupling** | Chirality transferred from spin edge states to lattice | Magnetoelastic coupling tensor, circular polarization selection, temperature | DCP, peak position and intensity contrast | Significantly asymmetric spectral weights | 100-130 cm$^{-1}$[29] |
| **Barnett rotation driving** | Mechanical angular velocity mapped to effective field | Rotation frequency, nonlinear strength | Oscillation threshold, second-order correlation | $\pm \boldsymbol{k}$ threshold difference, non-reciprocal blocking | Set by rotation and nonlinearity[99] |
| **Cavity-magnon-force tri-mode** | Coherent tri-mode amplifies directional selection | Tri-coupling amplitude and phase, device geometry | $S_{21}$ vs $S_{12}$ window and tuning | Programmable non-reciprocal window | MHz-GHz[98] |
| **Non-Hermitian exceptional points** | Complex eigenvalue merging and braiding | Gain, loss, asymmetric damping, encirclement path | Complex-frequency trajectories, final-state occupation | Encirclement chirality causes occupation flip, spectral braiding | Determined by linewidth and gain[91-93, 95] |

## 7 Summary and Outlook

Chiral magnons, as quasiparticle excitations in magnetically ordered systems of condensed matter, have evolved from empirical descriptions of dispersion asymmetry to a symmetry-centered materials physics framework. Their core lies in momentum inversion breaking, which endows spin waves with different eigenfrequencies and lifetimes at opposite wavevectors, thereby inducing directional selectivity in both the energy spectrum and dynamics. Altermagnetism provides a universal mechanism for momentum-dependent spin splitting in the weak-SOC non-relativistic limit, laying a realistic foundation for room-temperature, broadband candidate systems. The spin space group (SSG) places spin rotations on equal footing with crystal space groups, supplying symmetry criteria for chiral dispersion and nodal structures[94, 102]. These main threads have been mutually corroborated with characterization techniques such as INS, PNS, BLS, and THz spectroscopy, as well as first-principles calculations and LSWT, not only clarifying the intrinsic connections between material mechanisms and observables but also extending chirality from bulk non-reciprocity to system-level regulation involving many-body coherence and non-Hermitian topology.

At present, the challenges that urgently need to be overcome in this field focus on three aspects: room-temperature performance stability of materials, quantitative spectral-transport calibration, and many-body competition. Maintaining meV-scale splitting while suppressing linewidth at room temperature requires synergistic optimization through thin-film epitaxial engineering and low-damping oxide 2D van der Waals materials[80, 102]. Key parameters for realizing efficient room-temperature non-reciprocal transport include a Néel temperature $T_N > 300$ K to ensure thermal stability, an exchange constant $J > 10$ meV to amplify the amplitude of $\Delta\omega(k)$, and a damping parameter $\alpha < 0.01$ to extend the coherence length. These decisive metric thresholds can be locked onto candidate systems through high-throughput screening to avoid temperature-induced coherence decay. Mapping of $\Delta\omega(k)$ and $\Gamma(k)$ in crystal orientation and temperature range to SNE, SSE, and $\kappa_{xy}$ urgently requires simultaneous reporting of energy resolution and orientation information to achieve quantitative comparison[83]. Momentum and chirality dependence of multi-magnon scattering and damping kernels require precise microscopic characterization to lower the coherence threshold[51, 52].

To this end, future research in this field can develop synergistically in materials, theory and experimental characterization, and engineering realization. At the materials level, SSG classification combined with high-throughput computation can be used to preferentially screen multi-point-group altermagnets, establish databases labeled with splitting energy scales and lifetimes, and employ stress or hydrostatic pressure to amplify momentum selectivity and enter target frequency bands[58, 102]. On the theoretical side, it is necessary to go beyond conventional models and develop a unified spin-wave framework for two-magnon and multi-magnon processes that incorporates non-local and chiral damping, yielding experimental comparison phase diagrams[51, 95]. At the measurement level, quantitative description of sample crystal quality should be emphasized, and high-order processes should be systematically measured with platforms such as the China Spallation Neutron Source to verify spectral redistribution and linewidth anisotropy. In addition, at the engineering realization level, geometric phase engineering, as an important pathway for external non-reciprocal amplification, has demonstrated high feasibility. For example, in curved nanowaveguides, local geometric asymmetry introduces Berry phase shifts, enabling dynamic tuning of significant unidirectional transmittance in the GHz frequency range. Similarly, in artificial skyrmion lattices, topological charge reversal, combined with periodic potential-well design, can yield direction-selective magnon routing and broadband suppression.

Chiral magnon research pushes the multi-degree-of-freedom coupling of spin, lattice, and electromagnetic fields to a new frontier in condensed matter physics. Currently, symmetry-dominated materials design coordinated with cross-platform characterization has supported the verification of GHz-level prototypes. In the future, through many-body non-Hermitian expansion and room-temperature integration, breakthroughs in low-power consumption are anticipated in information processing, energy conversion, and quantum device fields[50, 53, 95].

The authors sincerely thank Xuan Guo for valuable suggestions during the writing of this review.